# Review

# Service Ecosystem: A Lens of Smart Society

Xiao Xue[1,*], ZhiYong Feng[1], ShiZhan Chen[1], ZhangBing Zhou[2], ChengZhi Qin[4,5], Bing Li[7], ZhongJie Wang[8], Bin Hu[9], ShuFang Wang[3], Hongyue Wu[1], Lu Zhang[1]

**Smart services are playing an increasingly important role in the operation of our society. Exploring the evolution mechanism, boundaries and challenges of service ecosystem is essential to our ability to realize smart society, reap its benefits and prevent potential risks. We argue that this necessitates a broad scientific research agenda to study service ecosystem that incorporates and expands upon the disciplines of computer science and includes insights from across the sciences. We firstly outline a set of research issues that are fundamental to this emerging field, and then explores the technical, social, legal and institutional challenges on the study of service ecosystem.**

Today, the emergence of intelligent technologies (such as the Internet, Internet of Things, big data, cloud computing, virtual reality, block-chain and artificial intelligence) are promoting the accelerated evolution of human society. The human society is like a container, in which these intelligent technologies continue to perform multi-form combinations and fusions to construct various intelligent applications[1-3]. In fact, "data + computing power + AI(Artificial Intelligence) algorithm = smart services"[4-6] is becoming a new type operating rules for the human society, so called smart society. As a complex social-technic system, its core challenge is to deliver high-quality municipal/social services at scale to an increased number of residents – despite limited capacities – with assistance from intelligent technologies.

The most significant difference between smart society and traditional society lies in the diversity of service species. In addition to human intelligent services popular in traditional societies, smart society also includes machine intelligent services and human-machine collaborative services[8]. On the one hand, the emergence of smart society is a bottom up process with self-organization and self-growth of various smart services. The state of the society is prescribed by the spontaneous interaction between human intelligence and machine intelligence. On the other hand, the smart society is a product following a top-down design. The rules revealed by machine intelligence with big data may change the planning and evolution path of human society.

The development and evolution of smart society involve many perspectives, including technology, society, economy, law, culture, etc. As the bridge between different elements and end users, "service" gives a unified description logic to everything in smart society, including applications, platforms, infrastructure, data, algorithms, knowledge, and everything else[4-6]. Furthermore, "service" plays the role of "connector" in the operation of smart society, making the collaboration and integration of cross-border elements truly possible[25-28]. Consequently, the entire society has gradually evolved to be an ecosystem of mutual dependence and mutual benefit, which are created and operated by different members in the society[7]. It is continuing to evolve driven by technological innovation, with energy and vitality beyond imagination.

In the process, the "chemical reaction" between smart services and human society drive those individuals, organizations, industries, and the world to be destructed granularly and restructured intelligently. At the level of daily life, the emergence of intelligent recommendation services (e.g. Amazon ECHO, Apple Homepod and AliGenie) has changed our daily behaviors and living habits to some extent, such as increased planning of behavior and decreased randomness or suddenness[9-11]. At the organizational level, intelligent technologies are changing traditional business mode in various fields. For example, the modes of education, finance, security, health, business, and social media are being redefined gradually[12-13]. At the level of social governance, the platform-based model has shown an increasingly obvious tendency to monopolize, while the non-governmental organizations (NGOs) and civil society groups are also becoming with more influential[1-3].

As the boundaries of service ecosystem in human society continue to expand, it will transform the way humans interact with their environment – in many cases, in ways that are not yet clear. Smart services have duality characterizes: sunny with a chance of thunderstorms. Although great progress has been achieved in many fields, smart services have also triggered a series of potential crises, such as the security of personal privacy[16,17], essential defects of machine intelligence[18-20], and potential risks of technological governance[21-24], and so on. Furthermore, the complex relationship between smart services exacerbates and magnifies these potential risks. Nowadays, it is too early to tell whether service ecosystem will evolve in the direction that people expect. The complexity often makes us fall into the dilemma of "to only see the tree but the forest" [14-15].

We need a new discipline that can integrate insights from inter-disciplines, which can reveal the full picture and evolutionary laws of smart society. The overview frames and surveys the emerging interdisciplinary field: service ecosystem. Here, we will outline the key research themes, questions and landmark research studies that exemplify this discipline. We starts by providing the background of service ecosystem and its interdisciplinary characteristics. We then provide a conceptual framework for the studies of service ecosystem. We close with a discussion of the technical, legal, economical and institutional barriers faced by researchers in this field.

[1]College of Intelligence and Computing, Tianjin University, Tianjin, China. [2]School of Information Engineering, China University of Geosciences, Beijing, China. [3]School of Geographic and Environmental Sciences, Tianjin Normal University, Tianjin, China. [4]State Key Laboratory of Resources & Environmental Information System, Institute of Geographic Sciences & Natural Resources Research, Chinese Academy of Sciences, Beijing, China. [5]College of Resources and Environment, University of Chinese Academy of Sciences, Beijing, China. [7]School of Computer Science, Wuhan University, Wuhan, China. [8]School of Computer Science, Harbin Institute of Technology, Harbin, China. [8]School of Information Science & Engineering, Lanzhou University, Lanzhou, China. * (Corresponding author: Xiao Xue，e-mail: jzxuexiao@tju.edu.cn）

## RESEARCH MOTIVATIONS

There are three primary motivations for the study of service ecosystem. First, various kinds of smart services are playing an ever-increasing role in our society, and service ecosystem emerges naturally during their interaction and evolution. Second, service ecosystem can provide a paradigm of describing smart society, and grasping its operation law is crucial to understand smart society. Third, because of its complexity properties, how to reveal the laws behind the evolution of service ecosystem poses a substantial challenge.

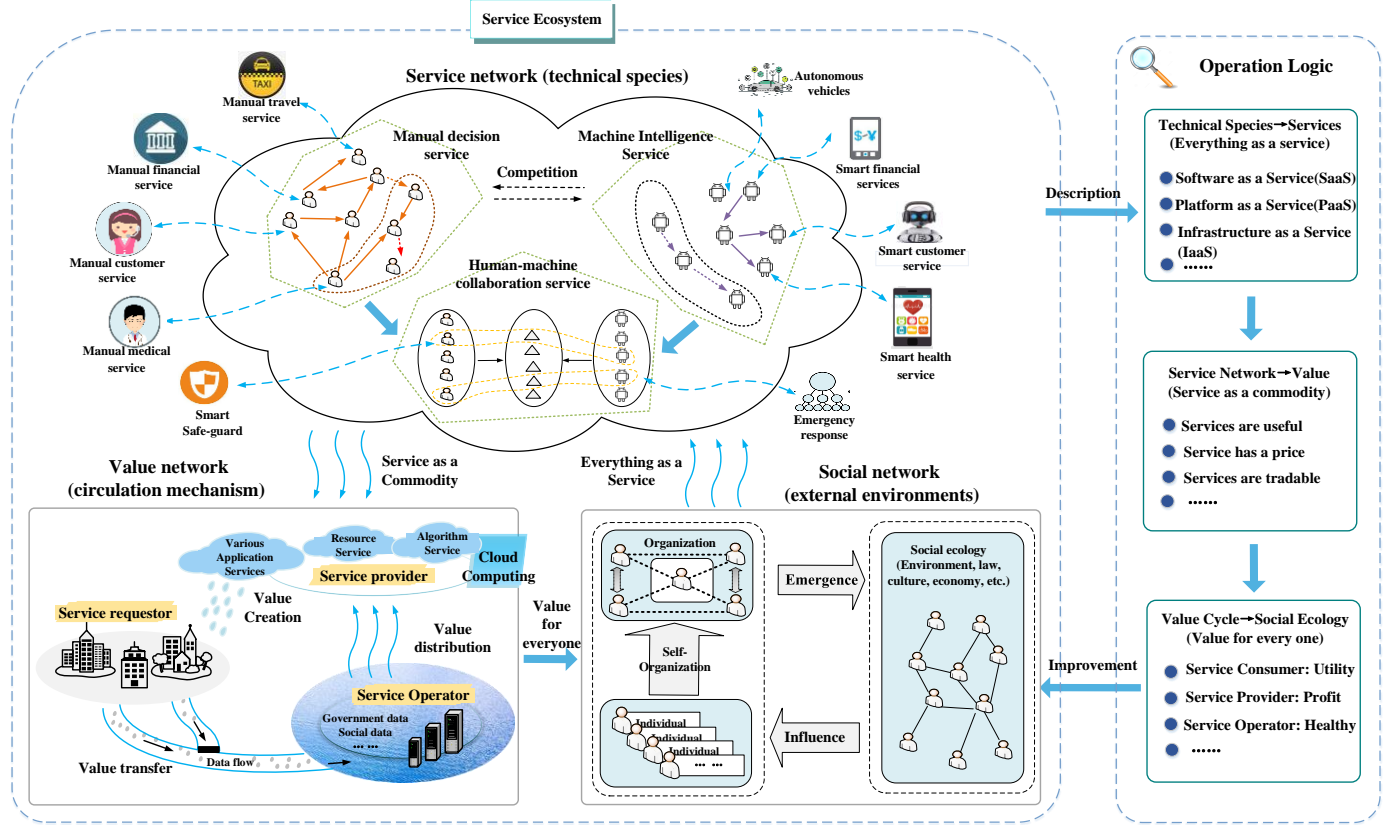

**Fig.1 | Operation diagram of service ecosystem.** Service ecosystem consists of three parts: (1) Service network (technical species): Various types of "service species" co-exist in cyber space, including human intelligence services, machine intelligence services, and human-machine coordination services. (2) Value network (circulation mechanism): As a transaction intermediary, service connects various roles (service demanders, providers and operators) and finally forms a value cycle[42]. (3) Social network (external environments): The social environments are an important factor influencing the implementation of smart services, including population, law, culture, customs and industries, etc., especially the coordination of interests among government, enterprises and citizens.

### The emergence of service ecosystem

The influence of smart service has already evidenced in all levels and domains of our society. The emergence and evolution of service ecosystem is mainly divided into three stages.

During the first stage (1998-2012), smart services were mainly applied to cyber space. To solve the information overloading problem in the Internet era, the recommendation services are constructed to bridge information resources and end users[9-11]. Amazon, Taobao, and so on are killer applications in the e-commerce field. Google, Netflix, Toutiao, and the equivalences are representatives in the information flow field.

During the second stage (2011-present), smart service began to spread in physical space. Traffic congestion, public service shortage (including finance, education and medical treatment), and environmental pollution all reflect the contradiction between limited service capacity and rapidly increasing demands. The goal of smart service is to either provide users with new alternatives or connecting them with new service resources, platforms and value. Uber, Meituan, Airbnb, etc. are typical representatives of O2O (Online to Offline) life services[12,13]; Alipay and PayPal are representatives of financial services; Alphabet's "Sidewalk Toronto" project[16,17], and Alibaba's "City Brain"[3] are representatives of smart cities.

During the third stage (2015-present), the continuous coupling of smart services and social space will eventually form a service ecosystem. This stage has the following characteristics: (1) Enhanced Integration. All human, things and environmental factors are all incorporated into service ecosystem, and are fully perceived, understood and calculated. (2) Comprehensive intelligence. Smart services would be widely distributed in society, seamlessly integrated with human intelligence. (3) Autonomous planning. In the presence of unknown risks, smart service can actively provide public services according to the situation, for example, the autonomous driving system of Apple and Google. If the above three goals can be achieved, some scholars even declared that it is possible to accurately predict the market, and ultimately achieve a planned economy[24].

### The operation diagram of service ecosystem

In the nature environment, all the things have their own niches and functions, meanwhile they are interwoven into a network. Driven by the continuous input of energy and material from external environments, the energy loop between species is formed through the food chain[63]. Finally, the natural ecosystem achieves a complex and dynamic balance, in which existing species may compete and die, while new species may appear and merge. Similar to natural ecosystem the operation diagram of service ecosystem is shown in Fig.1, which consists of three steps:

First, everything as a service (XaaS). All kinds of online or offline resources (applications, platforms, data, algorithms, and facilities) from social network are virtualized and published in the form of services[43-45]. Through their interconnection and cooperation, these services can realize the customization of a single resource and the on-demand aggregation of multiple resources[25-28]. Based on this, it is possible to create various "virtual organizations" (e.g. teams, enterprises and governments), as well as to redefine various public affairs (such as environment, traffic, education, health, etc.) [1-3].

Second, service as a commodity. The service matching between provision and demands can be finished in a "service marketplace", just like a commodity transaction, to achieve value creation, realization and distribution[42]. Enterprises will act as the service providers and give birth to smart services to satisfy ubiquitous demands. The public having diversified appeals may reach a consensus in a way, and bring about the emergence of demands. The governments have the responsibilities to guard social equality and public interests by virtue of policies, laws and regulations[16-17].

Third, value for everyone. Such a value circulation can promote the adjustment and evolution of social network, including the change of individual recognition and decision-making behaviors, cross-domain intersection and integration of organizations, and the refinement of the social governance mode. In the end, the value circulation will further promote the evolution of service network into a hierarchical structure similar to natural ecology, including individual, population, and community[7,46].

### The complexity of service ecosystem

According to our expectations, the sustainable development of service ecosystem not only takes into account the interests of all parties (citizens, enterprises, and government), but also needs to abide by a series of norms such as technology, economy, law, and ethics. However, as a complex social-technic system, it is hard to image what a service ecosystem will evolve into, and what we can recognize is certain sides only. Table 1 exemplifies those complexity characteristics and highlights those potential pitfalls in the development of service ecosystem. In order to change this situation, we need to be able to deal with the complexity challenges of service ecosystem, especially two key issues: (1)What is the laws behind the complex evolution phenomena of service ecosystem, especially when the application range of autonomous intelligent services is getting wider and wider? (2) How to identify the expansion boundary of service ecosystem to avoid its possible deviating from people's original intentions?

**TABLE.1 | Complexity Characteristics and Potential Pitfalls of Service Ecosystem**

| Complexity | Complexity Characteristics | Potential Pitfalls |
|---|---|---|
| **Two-way feedback:** | The two-way feedback mechanism (feedforward and feedback) exists between smart service and user data, which will produce a complex second-order emergence in service ecosystem[48]. As a result, the initial advantages tend to form a natural tendency to monopolize, and the possible adverse effects and potential pitfalls can be magnified fully[21-24]. | In terms of service consumers, users might suffer from information cocoons[29,30], algorithm discrimination[18-20], etc. As far as service providers are concerned, algorithm drift might lead to the inaccurate prediction in the case of emergencies, and platform economy might contribute to the rise of monopolies in industries, eventually stifling innovation and consumer choice [31-34]. |
| **Networked connection** | Because of the network characteristics, the assumption of individual independence in the system is no longer valid, which directly leads to various complex phenomena, such as nonlinearity, chaos, and emergence[47]. As a result, the trends of service ecosystem are uncertain and unpredictable. | In service ecosystem, local optimization of a single smart service does not necessarily lead to global optimization of the overall system[35-37]. For example, smart transportation services may alleviate traffic congestion in a certain local area, but due to the big data traps caused by limited resources and user games, it may have limited effect on alleviating global traffic conditions[38,39]. |
| **Critical phase transition** | With the emergence of a devastating technology or innovative business model, the service ecosystem may experience a state transition from one phase to another[49]. The failure to predict the financial crisis and stock market volatility are classic cases, where the reason lies in the continuous occurrence of various black swan events[40,41]. | An evolutionary game exists between the macroscopic regulation of social economic system and the autonomous behavior of individuals. It is difficult to get effective governance measures once and for ever, such as the change in urban planning[14] and conflicts in social ethical [37]. |

## INTERDISCIPLINARY RESEARCH

To study service ecosystem, especially its evolution and expansion, we must integrate knowledge from across a variety of scientific disciplines (Fig.2). This integration is currently in its nascent stages and has happened largely in an ad hoc domain in response to the growing need to understand service ecosystem. Currently, the scientists who most commonly study service ecosystem are the computer scientists, AI scientists and software engineers who have originally designed and created the smart services. These experts may be expert mathematicians and engineers; however, they are typically not trained complex system science. They often neglect the importance of social-systemic patterns of consumption and activity, in favor of a simplistic understanding of humans as atomized beings who make decisions based on economic and rational calculations. Conversely, sociological scholars have long been concerned about the political, economic and social dimensions of smart services and regarded it as one of the most typical and prominent technological governance representatives[21-24]. However, due to the non-transparency of smart services, as well as the lack of AI knowledge, their research have to be conducted in terms of "black box manner".

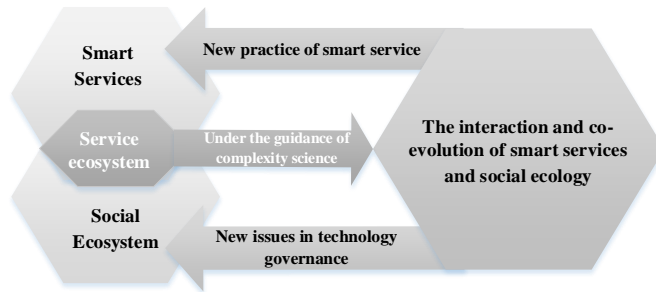

**Fig.2 | Multidisciplinary Research of Service Ecosystem.** Service ecosystem lies at the intersection of the fields that design and engineer smart services and the fields that traditionally use sociological methods to study human society. The insights from complexity science provide guidance that can help to study the laws of service ecosystem. In turn, those fields can provide new practice of smart services and identify the new issues of technological governance. Finally, service ecosystem provides a lens of smart society to make more precise statements about what smart society will be.

Integrating scientific practices from across multiple fields is not easy. So far, the main focus of those who create smart services has been on developing and expanding smart services in the corners of the entire society[50-53]. Excellent progress has been made in a wide variety of fields, ranging from face and speech recognition[54,55] to intelligent recommendation[9-11], autonomous locomotion[56], etc. Furthermore, these cross-domain services are connected to fulfill more complicated affairs, for example, smart healthcare, smart retail, smart public safe, and smart transportation, etc[25,26]. Finally, the scope of smart services covers the whole society, including building, communities, and cities[53,57]. In the construction of service ecosystem, varied service convergence technologies have been proposed, such as service composition (data convergence) [58,59], Big Service(capability convergence) [60], and crossover service(domain convergence) [61,62].

Under this context, researchers can use service ecosystem as a suitable research object to observe and analyze smart society, which has both technical and abstract characteristics. We hope to explore the potential effects of smart services on human society, define macroscopic or microscopic indicators to interpret the operation state of smart society, and predict whether intervention strategies can affect the evolution trend of smart society. In such a research process, other complementary fields can provide the inspiring thinking models, research tools, and a variety of optional conceptual frameworks.

### RESEARCH THEMES

DPSIR model is proposed by OECD (Organization for Economic Co-operation and Development) in 1993, which identified five complementary links of analysis that help to explain the relationship between social development and natural environment[64]. These links constitute a causal chain, i.e. *the driving forces -> pressure -> status -> impact -> responses*, which provides an organizing framework for studying the interactions between human behaviors and natural environment. Despite fundamental differences between natural ecosystem and service ecosystem, the study of service ecosystem can benefit from a similar analogy. Service ecosystem also has driving forces that promote evolution, undergoes operation that integrates social space into cyber space by means of value network, reaches a certain state under the combined effect of various factors, and embodies adaptive responses by continuously adjusting the relationship between smart services and social ecology.

Laura et al. believe that the DPSIR model can be used as a communication tool between environmental scientists and end users, but it underestimates the inherent uncertain and complex causality in environmental and socioeconomic systems[136]. In order to overcome the shortcomings of DPSIR model, we reconstructed the model to emphasize the unique characteristics of service ecosystem, which is summarized as five themes (DOSTR): *Driving forces, Operation, Status, Traceability, Response*. The goal of these distinctions is not division but rather integration. A complete understanding of service ecosystem will require integrating the five links. Scholars of computer science have already achieved substantial gains in understanding the whole life cycle of single smart service, although many questions remain. Relatively less emphasis has been placed on the evolution and intervention of service ecosystem composed of various services, i.e. the interaction between three heterogeneous networks (social network, service network and value network). We discuss these five topics in the next subsections and provide Fig. 3 as a summary.

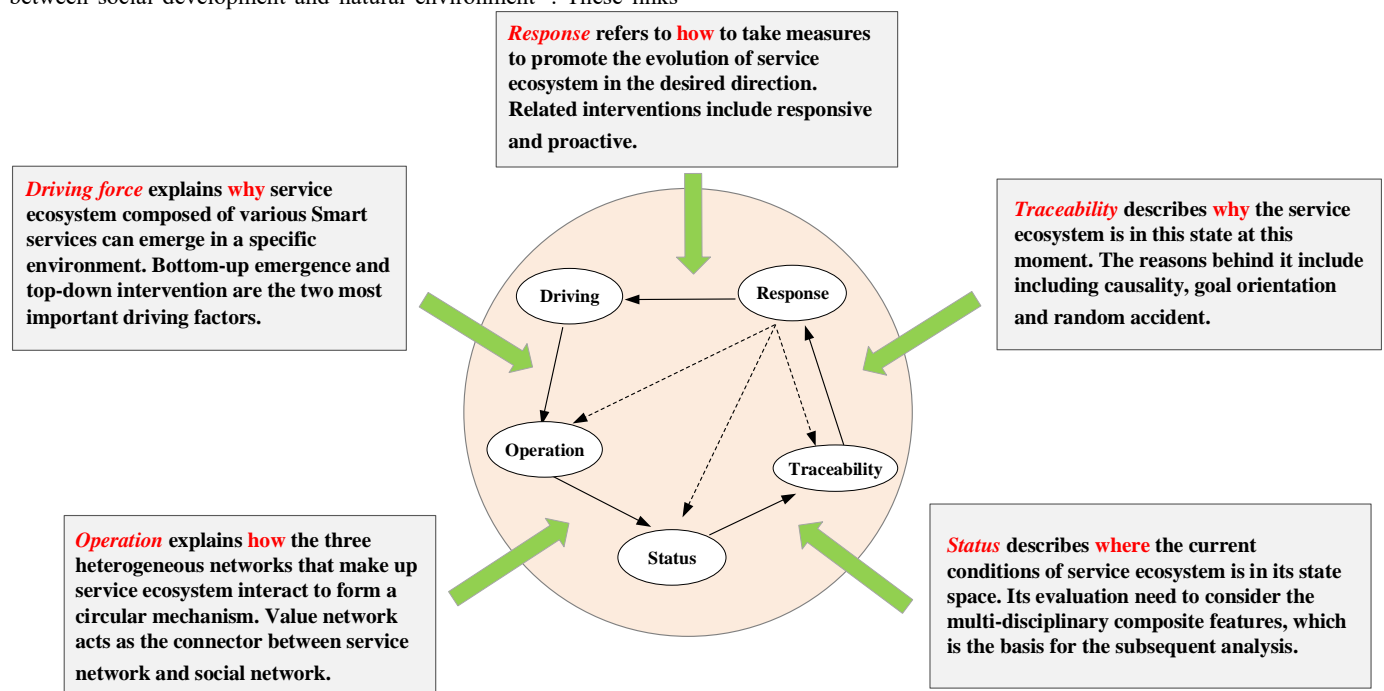

**Fig.3 | Research Theme of Service Ecosystem.** These five links reflect the logical chain of service ecosystem operation: *Driving force* is the premise, *Operation* represents dynamic characteristics, *Status* represents static characteristics, *Traceability* constructs the causality relations between up-to-date status and various factors, and *Response* means intervention and adjustment of service ecosystem.

#### Driving Forces

The operation and evolution of service ecosystem have to do with how the system behavior is observationally triggered and generated in specific environments. The bottom-up emergence and top-down intervention are the two most important driving factors.

The natural ecosystem relies on absorbing negative entropy to maintain its order[65]. In service ecosystem, the bottom-up emergence function as the negative entropy, such as the diffusion of technology innovation, the formation of fashion trends, and so on. They can promote the emergence of novel smart services, restart the coordination and optimization process of service resources, and reshape the growth curve of service ecosystem[66-68]. However, this kind of emergence is uncertain. It is hard to predict when and where they will occur, and their possible impact on the current status of service ecosystem. Currently, some technologies have shown potential that may trigger emergence, such as the 5G communication technologies, block-chain, and unmanned driving.

The top-down intervention is mainly to resolve the mismatch between supply and demand. However, the supply-demand matching is inherently uncertain. On the one hand, the initial state of service supply and demands are both uncertain, due to the sociality and diversity of their sources[69]. On the other hand, the matching status of service supply and demand often change dynamically during the long-lasting process. –In this case, the intervention strategies are often carried out from two aspects: improvement of matching efficiency (e.g. leveraging data gathered through video feeds at traffic lights to ease traffic congestion) and enhancement of service capabilities (e.g. using intelligent technologies to expand the scope of existing financial services to uncover new needs).

A bilateral circulation exists between these two driving factors. The bottom-up emergence can help to redefine current social procedures (planning, construction, operation and management) with an interactive, sharing, elastic and competitive model. However, this improvement may solve old problems, but introduce new ones. In order to meet new challenges, the top-down intervention needs to promote the emergence of novel smart services as well as the optimization and upgrading of existing services. Sharing economy is a typical example of two driving forces working together, which can make full use of mobile platform to bridge the gap between those originally idle service resources and the widespread public demands (bottom-up)[70,71]. But, the service quality and user safety issues arising from this novel services need to be regulated by new laws (top-down).

*Operation*

In the study of service ecosystem, operation describes how a value cycle contributes to the co-evolution of service network (technical species) and social network (external environment). The current niche of each service is gradually formed in the long-term operation of service ecosystem based on the survival of the fittest.

Social network focuses on the impact of human behavior on smart services, which consists of service consumers, service providers, service operators and their groups. Through self-growth and self-organization of these roles, the system-level trends can emerge and then influence the evolution trends of smart services. The behavioral data of service consumers is the raw material of smart services, which can affect the performance of smart services in a passive way. Service providers can create novel smart service or improve –service performance in a positive way. By analyzing the conditions of supply and demand matching, the system level service operation strategy can be adjusted, which can influence the competition between services in turn. Such an evolution phenomenon is a case of the 2$^{nd}$ order emergence[72].

Service network focuses on the utility value of smart services for humans, which is composed of all available services and their interactive relationships. The territory of service network is gradually formed in the competition and cooperation of services. Smart services constantly explore new opportunities in traditional domains, and gradually gain a competitive advantage over manual services[73]. By providing users with more useful services, service provider can get more revenue and benefit a number of parties. Service operators can expect more efficient use of resources, and service consumers can get a better service experience.

The value network acts as a connector between service network and social network, which represents the value in service ecosystem, similar to the energy circulation in natural ecosystem[29,74]. First, by means of the service supply and demand matching, the value creation is realized of services can be realized through satisfying users' needs. Second, the user pays for the service, which realizes the value delivery between service consumers and service operators. Finally, service operator implements the secondary distribution of value in related service providers. Different service operation strategies will lead to great differences in the operation efficiency of value cycle, which in turn affects the adjustment of social networks.

*State*

In the study of service ecosystem, state refers to how to identify the position of the system in its state space from different perspectives. Its measurement and evaluation need to consider the multi-disciplinary composite features, such as ecological system, complex network, and social-economic system.

The ecological status are mainly used for macro-evaluation of service ecosystem, which are developed from three dimensions, i.e. external environment, ecological species and circulation mechanism. In the environmental dimension, the research is mainly on how the system obtains negative entropy from the environment to maintain the order and output of the system. The indicators include openness, sustainability, etc[75]. In the species dimension, the research is mainly on the competition of different species for ecological niche in the system, and the indicators include diversity, hierarchy, etc. In the cycle dimension, the research is mainly on the features and changes of the system at different stages with focus on future forecasts. The indicators include periodicity, anti-interference, etc[76].

The network characteristics are mainly applied for micro-measurement of service ecosystems, and its goal is to construct a relationship map of service species[77,78]. The commonly used approaches consists of the following three categories: (1) The relational approach focuses on the edges between nodes, which are used to describe specific behaviors and processes. The indicators applied include density, scale, centrality, symmetry, etc. (2) The positional approach targets at the network structure constructed by all nodes, and emphasizes the importance of structural equivalence in understanding social behavior. The indicators used include community, hot spots, structural voids, path length, etc[79,80]. (3) The performance approach studies the performance of the entire network system. The indexes include functionality, responsiveness, reliability, availability etc.

The economic viability is critical to the long-term success of service ecosystem. Even the most novel idea with enormous potential to improve social operations may never deliver on its promise if it proves economically unsustainable (for unforeseen reasons). Here, researchers have developed considerable interest in the three related laws: (1) Metcalfe's Law, that is, the value of network economy is equal to the square of the number of network nodes. The power of service ecosystem will grow exponentially along with the increase of smart services[81,82]. (2) Matthew Effect, that is, "the strong being stronger and the weak being weaker". The current development of smart services depends on the application of big data, having strong economy of scale and economy of scope. This will lead to a serious divergence in the number of users of different services in the service ecosystem[83,84]. (3) Marginal Utility: The service cost will decrease progressively with the increase of the system scale. But after the scale reaches to a certain level, the invest and produce will approach to the bottleneck of accumulation and growth. Service ecosystem cannot expand without limitation, due to the constraint of various policies, resources and technologies[85-87].

*Traceability*

In the study of service ecosystem, traceability helps us to understand why some smart services spread and persist while others decline and vanish. In addition to cause determinism, its evolution paths is affected by people's initiative and random accidents.

The cause determinism is largely dependent on the digital mapping relationship between cyber world and real world. When there are various problems and contradictions in the real world itself, the cyber world will also present the same problems. In many cases, the positive feedback characteristics of machine intelligence will strengthen and amplify problems, such as information cocoon[34,35] and algorithmic discrimination[18-20]. In addition, whether the mapping relationship is complete and accurate will directly affect the performance of smart services, which may induce potential pitfalls, such as the lack of ability to deal with emergencies[88] and the "disastrous forgetting" in cross-domain applications[89,90]. With the increasing complexity of system, it is becoming more and more difficult to find the causality[91-93], and sometimes we have to use strong correlation instead[94,95].

The goal orientation reflects people's free will and bounded rationality. Both manual decision-making services and machine intelligence services are purposive and initiative. They can continuously improve their "survival skills" through mutual learning and interaction with the environment, thus to survive and evolve. Such purposive behaviors of services may result in unfair competition and undermine the healthy development of service ecosystem, such as a bidding ranking of search engines, the brush reputation of e-commerce etc[96]. The stable organizational behavior patterns will emerge from a large number of individual autonomous behaviors, for example, the scale-free effect caused by preferential attachment[97]. The purposive interaction between individuals and the environment will cause the ecosystem to constantly reshape its own evolution direction, strategy and structure.

The random accident refers to those sudden and uncertain events beyond our prediction capability, e.g. suddenly popular people and events in the online social space[98,99]. In service ecosystem, the source of service provision and service demands are social, which greatly increases the probability of accidental events. Fundamentally speaking, many problems to be solved by smart services are actually the problems of how to eliminate uncertainty. However, the information we own is always little in comparison with the complexity of problems, which often leads to the failure of current intelligent technologies. In order to change the situation, the following thinking models need to be emphasized: evolutionary algorithm (trial and error instead of design, redundancy response to fragility), reinforcement learning (clear goals, constantly adjusting between exploration and benefits), and Bayesian model (adjust your beliefs based on facts, but not too quickly) [100,101].

*Response*

The evolution of service ecosystem may follow different trends, either evolving from low level to high level; or degrading from high level to low level. Response refers to how to adjust the evolution direction of service ecosystem through conducting some reasonable and limited interventions.

The logic of service ecosystem is actually to transform the urban space into a Cyber-Physical-System (CPS) using Information & Communication Technologies (ICT). Such a renovation follows the basic logic of cybernetics, which is to identify the difference between the practical effect and expected goal, and then take correction measures to promote the system to the expected state through cyclic feedback[102]. By virtue of the large scale collection and analysis of observational and statistical data, the mismatching or overfitting between service supply and demand can be discovered timely, which can provides the possibility for the refined and effective intervention. After promptly identifying the correlation among various factors and problems, decision makers can conduct the proper reactive responses, iteratively detect feedback then to adjust their behaviors, and even make proactive prevention before the appearance of problems.

Reactive response is mainly aimed at existing services and promote the improvement of services through their social utility evaluation[103,104], such as the routes re-planning of navigation services to avoid roads with high congestion. Proactive response is mainly for newborn services, and occupy new ecological niches by discovering structural holes in the network[78], such as Alibaba's advance layout in cloud computing infrastructure services. Compared with reactive responses, the proactive approach tries to prevent concerning or predefined events from taking place. Advanced data predictive tools allow service operators to foresee future events, which is developed by combining regular patterns mined from historical data and irregular patterns mined from live data (that is, the most fresh and recent data).

Currently, many reactive responses can be conducted by autonomous intelligence services. The operation of our society will become more efficient by rationally positioning smart services in service ecosystem and precisely setting their behavior rules[105,106]. Web-robots of Wikipedia and robot assistants of social network[107] are typical examples of autonomous reactive response by smart services. However, proactive responses often need to deal with more complex scenarios, which require a coordination across multiple organizations or industries, and the adaptability to constraints caused by various potential changes. In this situation, the man-machine coordination services might be more suitable. Some successful practices with this model have been developed, such as, crowdsourcing in human computation[108,109], and group perception in urban computing[110-112].

### RESEARCH SCALES

With the framework outlined above and in Fig. 3, we now catalogue the effect of service ecosystem at the three scales of inquiry: individual behavior and cognitive ability (micro-scale), competitiveness of the organization or institution (medium-scale), and the governance capacity of the entire society (macro-scale). Here we emphasizes the function of single smart service itself, the collective capability of multiple services, and the complex characteristics of service ecosystem. Fig.4 shows a matrix for evaluating the impact of service ecosystem, which address the possible questions encountered in different links (Fig.3): operation/status, analysis/traceability and intervention/adjustment.

| Impact of Service Ecosystem / Scales of Inquiry | Operation/Status | Analysis/traceability | Intervention/adjustment |
|---|---|---|---|
| Micro-level (Individual behavior) | ✓ | ✓ | ✓ |
| Medium-level (Organizational competitiveness) | ✓ | ✓ | ? |
| Macro-level (Social governance ability) | ✓ | ? | ? |

**Fig.4 | Research Scales of Service Ecosystem.** In this matrix, options with a check mark represents the fields where smart services do have a lot of achievements, and options with a question mark represents the fields where smart services only have limited achievements but requires a careful evaluation.

*Micro-scale*

The micro-scale study of service ecosystem focuses on the impact of smart services on individual behavior and cognitive ability. Often these studies focus on the formation of a service community with individualization as the goal. The fields of machine learning and software engineering currently conduct the majority of these studies. For example, the popular personal intelligent assistant services largely depends on the aggregation of recommendation services in different fields, such as purchasing, investment, healthcare, travel, and so on[117,118]. Furthermore, smart services can subtly influence individual behavioral decisions from multiple channels, including personal recommendation, circle of friends, official accounts, hot searches, etc.

To make smart services more effective, it is necessary to construct the electronic archive of "digital self". Thanks for the sustainable development of sensing and communication technology, the interactions between people and cyber-physical worlds can be digitalized, recorded, and collected, opening up new opportunities for monitoring urban activity at an unprecedented scale[113]. For instance, wearable sensors record a person's physiological parameters (e.g., heart rate or blood sugar level); a user's activity on online social services can be easily tracked from the generated content (e.g. event logs, twitter data) [114-116]. Through the data analysis and data mining, the personalized information (e.g. consumption patterns, lifestyle, aesthetics, and even value orientation, etc.) of people can be acquired.

The convenience of smart life does conflict with the risk of manipulating individual cognition. How to develop effective operation models to balance the two has become the focus of future research[8,119]. The collaboration, competition and evolution of service ecosystem are the factors which potentially impact individual cognition, but have been poorly modeled until now. In Social Learning Theory proposed by Albert Bandura, it is confirmed that Human intellectual development relies on individual's innate foundation, the acquired learning environment and sociocultural influence. In a word, the formation and change of individual cognition is the result jointly affected by individual learning, organizational learning and cultural influence[72,120]. This viewpoint provides ideas on how to analyze the impact of service ecosystem on individuals.

*Medium-scale*

The medium-scale study of service ecosystem focuses on the effect of smart services on traditional domains or existing organizations. Being both powerful and relatively cheap, smart services will spread faster than computers did and touch every industry. Recruiters are able to pinpoint the best candidates more easily, and customer-service staff are able to handle queries faster. Jobs that never existed before could be created. The radical changes are undergoing in industries like health care and transport that could lead to new drug discoveries and treatments and safer ways to move around. In the territory of traditional industries, the novel smart services are constantly growing, expanding, and disruptive, and meanwhile many traditional services are shrinking, merging and even dying out in competition.

We hope to find out the competitive advantages of smart services over traditional modes, which determines whether smart services can disrupt traditional industries. It can be analyzed from the point, line, plane and body dimensions. Wherein, points are composed of service consumers and service providers. Lines connect the users and the services, which reflect service invocation relations in applications. When service points are interlinked with each other to provide an integrated service model, the plane will be derived. The one point (user or service) may locate in multiple planes. When the number of planes increases, these intertwined planes eventually constitute the body, i.e. the domain service ecosystem.

Today many firms are competing to constructing their domain ecosystem by means of smart services. In this way, they can obtain the greatest customer stickiness and earn more profits[13]. This has been proved by the emerging Internet giants, for example, Uber, Airbnb, Meituan, etc[121-123]. However, the nature of production, the availability of data, the scope of changing business processes and the industrial structure of different industries are completely different, such as the health industry and the financial industry. This means that smart services will have completely different impacts on different industries[13]. This would be of great concern in the near future.

*Macro-scale*

The macro-scale study of service ecosystem focuses on the impact of smart services on social governance capability, e.g. the reduction of market disorder and promotion of planned intervention. From the perspectives of CPS or controllability, the society can be composed of natural ecological environment (explainable is our goal, e.g. the discovery and traceability of the sudden pollution incidents), man-made environment (completely controllable is our goal, including various engineering infrastructures and buildings, landscapes, streets, etc.), and crowd behavior (enhanced planning is our goal, including business, life, production, etc.)[124,125]. For the first two types of systems, human activities are only as their input variables. The last type of system is built around people's demands and behaviors. At present, the majority of smart society projects falls in the third category, such as education, medical care, retail, tourism, and government affairs.

However, the randomness and freedom degree of crowd behavior result in the fact that smart services has weak capacities of perception and intervention at the macro scale. In this situation, those centralized control and optimization methods are simple, identical and average, which are hardly cope with various dynamics, diversity and uncertain scenes. Currently, the governance model of service ecosystem can be generally divided into three categories: the market approach based on the transaction cost theory, the control approach based on hierarchical system, and the governance approach between the two[126,127].

According to the Law of Requisite Variety[128,129], the effective social governance model must be "specific" to each individual in the environment, in order to respond to all possible scenarios in the environment. However, such a model cannot be acceptable in cost. Otherwise, an improvement of governance efficiency at a macro scale means a loss of individual autonomy at a micro scale. For example, the collaboration efficiency can be improved by reducing the competition between single services, but with the loss of system adaptability. Therefore, we need to pay more attention to how to implement intervention at a reasonable scale, rather than intervening in the autonomic behavior of each individual. Note that, the performance of varied intervention strategies may vary significantly due to different implementation details. It is a feasible way to deduce and evaluate their performance in "artificial laboratories" before being put into practice, which could minimize the cost of trial and error[130-132].

## OUTLOOK

Whether we like it or not, our society is being fully affected, transformed and shaped by ubiquitous, transparent and personalized services. Only by capturing their evolutionary laws and expansion boundaries of smart services in human society, can we make reasonable and limited interventions. In this setting, we need a new interdisciplinary research field: Service Ecosystem. For the smooth development of this descipline, the following issues are to be taken special attention.

First of all, service ecosystem involves in a dynamic evolution, with novel services to be emerging continuously. These novel technological species will affect the current operating mode of society in various ways. As the diagram of smart society, service ecosystem needs to identify and predict the impact of various new smart services on existing social systems, so that our understanding of smart society can keep pace with times.

Secondly, most of scientists put emphasis just on technical aspects of smart services, leaving apart social elements involved in it. Due to the huge differences in social systems and local cultures, the performance of smart services varies greatly in different environments. Therefore, researchers shall pay much attention to the influence of localization characteristics on the evolution of service ecosystem. The perfect working process might lie somewhere between local characteristics and general regulations.

Thirdly, in the construction of smart services, a low level of investment will be insufficient, but a high level of investment may get its own benefit impaired a lot when it improves users' utility considerably. Therefore, a critical research issue of service ecosystem is how to find a delicate balance between social affordability and expected benefits. No matter how ambitious the target is, it should follow the principle of gradual improvement to prevent the ratio of earnings to costs from continuously decreasing.

Fourthly, it is too early to tell whether the positive changes wrought by smart services will outweigh the perils. The essence of evolution is to constantly produce random "errors" in response to unforeseen situations. The elimination of uncertainty by smart services will enable the entire society to achieve great development in a short period of time, but it may be at the expense of future changes and innovation capabilities. Therefore, the research on service ecosystem has to pay much attention to how to implement interventions of appropriate scales to avoid overkill.

Fifthly, the research on service ecosystem always requires the mapping relationships between physical space and virtual space to be accurate, comprehensive and dynamic. Even the trajectory of entity objects should also be included for the research. Because these "social laboratories" may bring privacy violations, the ethical considerations should be under strict supervision and legal restraint.

Finally, the exploration in this disciplinewill require the joint efforts from those related disciplines, because these studies are accompanied by challenges brought about by interdisciplinary cooperation. It is essential to meet these challenges. Universities, governments, and funding agencies should play an important role in developing a large-scale, equal and credible interdisciplinary research.

To summarize, smart services are promoting the refined social governance to an unprecedented level, which can facilitate our life significantly and improve the utilization of social resources to a large extent. Studying service ecosystem can support us in examining technological and environmental sustainability dimensions jointly with social justice perspective.

RESEARCH REVIEW

**Acknowledgements** Thanks for the support provided by National Key Research and Development Program of China (No. 2017YFB1401200), National Natural Science Foundation of China (No.61972276, No. 61832014, No. 41701133).